# A Robust Ranging Scheme for OFDMA-Based Networks

Michele Morelli, *Senior Member, IEEE*, Luca Sanguinetti, *Member, IEEE*, and H. Vincent Poor, *Fellow, IEEE*.

## Abstract

Uplink synchronization in orthogonal frequency-division multiple-access (OFDMA) systems is a challenging task. In IEEE 802.16-based networks, users that intend to establish a communication link with the base station must go through a synchronization procedure called Initial Ranging (IR). Existing IR schemes aim at estimating the timing offsets and power levels of ranging subscriber stations (RSSs) without considering possible frequency misalignments between the received uplink signals and the base station local reference. In this work, we present a novel IR scheme for OFDMA systems where carrier frequency offsets, timing errors and power levels are estimated for all RSSs in a decoupled fashion. The proposed frequency estimator is based on a subspace decomposition approach, while timing recovery is accomplished by measuring the phase shift between the users' channel responses over adjacent subcarriers. Computer simulations are employed to assess the effectiveness of the proposed solution and to make comparisons with existing alternatives.

M. Morelli and L. Sanguinetti are with the University of Pisa, Department of Information Engineering, Via Caruso 56126 Pisa, Italy (e-mail: michele.morelli@iet.unipi.it, luca.sanguinetti@iet.unipi.it). This work was completed while L. Sanguinetti was with Princeton University and it was supported by the U.S. National Science Foundation under Grants ANI-03-38807 and CNS-06-25637. This paper was presented in part at the IEEE International Conference on Communications (ICC), Beijing, China, 2008.

H. Vincent Poor is with the Department of Electrical Engineering, Princeton University, Princeton, NJ 08544, USA (e-mail: poor@princeton.edu)

# I. INTRODUCTION

The demand for high data rates in wireless communications has led to a strong interest in multicarrier modulation techniques, and particularly in orthogonal frequency-division multiple-access (OFDMA), which has become part of the IEEE 802.16 family of standards for wireless metropolitan area networks (WMANs) [1].

Despite its many appealing features, OFDMA is extremely sensitive to timing errors and carrier frequency offsets (CFOs). The former give rise to interblock interference (IBI), while the latter produce interchannel interference (ICI) as well as multiple access interference (MAI). To cope with such impairments, the IEEE 802.16 standards specify a synchronization procedure called Initial Ranging (IR) by which users adjust their transmission parameters so that uplink signals arrive at the base station (BS) synchronously and with approximately the same power level. In its basic form, the IR process develops through the following steps. First of all, each ranging subscriber station (RSS) computes frequency and timing estimates on the basis of a downlink control channel. The estimated parameters are used in the subsequent uplink phase, during which each RSS transmits a randomly chosen code over a ranging time-slot. As a consequence of the different users' positions within the cell, uplink signals arrive at the BS at different time instants. Furthermore, since the ranging code is randomly selected, several users may collide over a same time-slot. After identifying colliding codes and extracting timing and power information, the BS will broadcast a response message indicating which codes have been detected and giving instructions for timing and power adjustment.

From the above discussion, the main functions of the BS during the ranging process may be classified as multiuser code detection and multiuser timing/power estimation. Some methods to accomplish these tasks were originally suggested in [2] and [3]. In these works, a long pseudo-noise (PN) code is transmitted by each RSS over all available ranging subcarriers. Code detection and timing recovery is then accomplished on the basis of suitable correlations computed in either the frequency or time domains. This approach requires huge computational complexity since one correlation must be evaluated for each possible ranging code and hypothesized timing offset. Moreover, in the presence of multipath distortions ranging subcarriers are subject to different attenuations and phase shifts, thereby leading to a loss of the code orthogonality. This gives rise to MAI, which severely degrades the system performance. Alternative solutions can be found in [4] and [5]. In particular, the method in [4] replaces the PN ranging codes with a set of modified generalized chirp-like (GCL) sequences and mitigates the effects of channel distortion through differential detection of the ranging signals. Unfortunately, this approach is still plagued by significant MAI. The scheme discussed in [5] aims at reducing the system complexity by breaking the multiparameter estimation problem into a number of successive steps. However, it is specifically devised for flat channels and fails in the presence of multipath distortions.

All previously discussed methods exhibit poor performance in the presence of frequency selective channels. Such drawback is partly mitigated in [6] by employing ranging subchannels composed by a small set of adjacent subcarriers over which the channel gains do not vary significantly. In order to achieve multiuser and multiantenna diversity gains, each RSS exploits channel estimates obtained during the downlink phase to select the best subchannel (i.e., the one characterized by the largest power gain). Code detection is then accomplished by correlating the received frequency-domain samples with the corresponding code sequence and comparing the result with a pre-assigned threshold. In the presence of timing offsets, however, the received codes are affected by different linear phase shifts and loose their orthogonality, thereby leading to residual MAI. A signal design which is robust to multipath distortions is proposed in [7], where ranging signals are divided into several groups and each group is transmitted over an exclusive set of subcarriers with a specific timing delay. This approach leads to a significant reduction of MAI as signals of different groups are perfectly separable in either the frequency or time domain. Better results are obtained with the orthogonal signal design presented in [8]. In this scheme, each RSS selects a ranging subchannel composed of a specified number of subcarriers and transmits a randomly chosen code over adjacent OFDMA symbols. Spreading is thus performed in the time domain as the same code is transmitted in parallel over all selected subcarriers. In a perfectly frequency synchronized scenario, codes transmitted on different subcarriers remain disjoint at the receiver. Furthermore, if the channel response keeps constant during the overall ranging period, codes received over the same subcarrier are still orthogonal and can easily be separated at the BS. After multiuser code detection, timing information is eventually acquired in [8] through an iterative procedure, which exploits the autocorrelation properties induced by the cyclic prefix (CP). In spite of the improved robustness against channel distortions, spreading across adjacent symbols increases the sensitivity of the system to residual CFOs. In [8] it is assumed that during the ranging process frequency errors are so small that the demodulated signals incur only negligible phase rotations over one OFDMA symbol. However, phase rotations may become significant if the ranging period spans several adjacent symbols. In such a case, the received ranging signals are no longer orthogonal and CFO compensation is necessary to avoid a serious degradation of the system performance.

In the present work we adopt the orthogonal signal design of [8] and propose a novel ranging method for OFDMA networks that is robust to frequency errors. To avoid complex multidimensional optimizations, we adopt a multistage approach where the number of active codes is first determined by resorting to the minimum description length (MDL) principle [9], and the multiple signal classification (MUSIC) [10] algorithm is next employed for code identification and CFO estimation. Frequency estimates are then used in the third step, where timing and power level estimation is accomplished in an ad-hoc fashion.

The rest of the paper is organized as follows. Section II describes the adopted signal model. In Section III we address the problem of code identification and CFO recovery, while Section IV is devoted to timing and power estimation. Section V illustrates a method to detect possible collisions between RSSs that use the same code and ranging subchannel. Simulation results are presented in Section VI and some conclusions are drawn in Section VII.

*Notation*: Matrices and vectors are denoted by boldface letters, with $\mathbf{I}_N$ and $\mathbf{0}_N$ being the identity and null matrices of order $N$, respectively. $\mathbf{A} = \mathrm{diag}\{a(n)\,;\; n = 1, 2, \ldots, N\}$ denotes an $N \times N$ diagonal matrix with entries $a(n)$ along its main diagonal, while $\mathbf{B}^{-1}$ and $\mathrm{tr}\{\mathbf{B}\}$ are the inverse and the trace of a square matrix $\mathbf{B}$. We use $\mathrm{E}\{\cdot\}$, $(\cdot)^*$, $(\cdot)^T$ and $(\cdot)^H$ for expectation, complex conjugation, transposition and Hermitian transposition, respectively. The notation $\|\cdot\|$ represents the Euclidean norm of the enclosed vector while $|\cdot|$ stands for the modulus. Finally, $\mathrm{round}\{x\}$ indicates the integer closest to $x$ and $[\cdot]_{k,\ell}$ denotes the $(k, \ell)$th entry of the enclosed matrix.

## II. System description and signal model

### A. System description

The investigated OFDMA system employs $N$ subcarriers with frequency spacing $\Delta f$ and indices in the set $\mathcal{J} = \{0, 1, \ldots, N-1\}$. To avoid aliasing problem at the receiver, a number $N_0$ of null subcarriers are placed at both edges of the signal spectrum. We denote by $R$ the number of ranging subchannels and assume that each of them is divided into $Q$ subbands uniformly spaced over the signal bandwidth at a distance $(N_U/Q)\Delta f$ from each other, where $N_U = N - 2N_0$ is the number of modulated subcarriers. A given subband is called a *tile* and is composed by a set of $V$ adjacent subcarriers. The subcarrier indices in the $q$th tile $(q = 0, 1, \ldots, Q-1)$ of the $r$th subchannel $(r = 0, 1, \ldots, R-1)$ are collected into a set $\mathcal{J}_q^{(r)} = \{i_{q,\nu}^{(r)}\}_{\nu=0}^{V-1}$ with entries

$$i_{q,\nu}^{(r)} = \frac{qN_U}{Q} + \frac{rN_U}{QR} + N_0 + \nu. \tag{1}$$

The $r$th subchannel is thus composed of subcarriers with indices taken from $\mathcal{J}^{(r)} = \cup_{q=0}^{Q-1} \mathcal{J}_q^{(r)}$, while a total of $N_R = QVR$ ranging subcarriers are available with indices in the set $\mathcal{J}_R = \cup_{r=0}^{R-1} \mathcal{J}^{(r)}$. The remaining $N_D = N_U - N_R$ subcarriers are used for data transmission and are assigned to data subscriber stations (DSSs) which have successfully completed their IR process at an earlier stage. We denote by $M$ the number of OFDM symbols in a ranging time-slot. In the sequel, we assume that $M$ is a power of two.

The proposed ranging process develops through the following steps:

1) Each RSS selects one of the $R$ available ranging subchannels according to some specified criterion. In low mobility applications, channel estimates obtained during the downlink slot can be exploited to choose the less attenuated subchannel [6]. Here, we follow a simpler approach in which the ranging subchannel is randomly selected by each RSS.

2) The RSS transmits a randomly chosen code of length $M$ during the ranging time-slot. Similarly to [8], such a code is transmitted in parallel over all subcarriers belonging to the selected subchannel and is taken from an orthogonal set $\mathcal{C} = \{\mathbf{c}_1, \mathbf{c}_2, \ldots, \mathbf{c}_M\}$ (e.g., a Walsh-Hadamard set) with $\mathbf{c}_k = [c_k(0), c_k(1), \ldots, c_k(M-1)]^T$ and $|c_k(m)| = 1$ for $0 \leq m \leq M-1$.

3) On the basis of the received uplink signals, the BS determines which codes are actually being employed and extracts the corresponding frequency, timing and power information. It also detects possible collisions between RSSs that use the same code and ranging subchannel.

4) Once the above operations have been successfuly completed, the BS will broadcast a response message by which the detected RSSs can adjust their synchronization parameters. Since users that do not find their ranging information in the response message must re-initiate the ranging procedure in the next frame, there is no need to notify the collision status to collided RSSs.

Without loss of generality, in the ensuing discussion we concentrate on the $r$th subchannel and assume that it has been selected by $K^{(r)}$ RSSs. To simplify the notation, the subchannel index $^{(r)}$ is dropped henceforth.

The waveform transmitted by the $k$th RSS propagates through a multipath channel with impulse response $\mathbf{h}_k = [h_k(0), h_k(1), \ldots, h_k(L_k - 1)]^T$. As the channel order $L_k$ is usually unknown, in practice we replace $\mathbf{h}_k$ by an $L$-dimensional vector $\mathbf{h}'_k = [\mathbf{h}_k^T, 0, \ldots, 0]^T$, where $L \geq \max_k \{L_k\}$ is a design parameter that depends on the maximum expected channel delay spread. At the BS, the received samples are not synchronized with the local references. We denote by $\theta_k$ the timing error of the $k$th RSS expressed in sampling intervals while $\varepsilon_k$ denotes the frequency offset normalized by the subcarrier spacing. As explained in [11], users that intend to access the network compute initial frequency and timing estimates on the basis of a downlink control signal broadcast by the BS. The estimated parameters are then employed by each RSS as synchronization references for the uplink ranging transmission. This means that during IR the CFOs are only induced by Doppler shifts and/or downlink estimation errors and, in consequence, they will be quite small. As an example, consider the IEEE 802.16 standard for WMANs with subcarrier spacing $\Delta f = 11.16$ kHz and carrier frequency $f_0 = 2.5$ GHz. In case of ideal downlink synchronization, the maximum CFO in the uplink is $2f_o v_k / c$, where $v_k$ denotes the terminal speed while $c = 3 \times 10^8$ m/s is the speed of light. Letting $v_k = 120$ km/h yields $\varepsilon_k \leq 0.05$, which means that the normalized CFO lies within 5% of the subcarrier spacing. Timing errors depend on the distances of the RSSs from the BS and their maximum value corresponds to the round trip propagation delay for a user located at the cell boundary [11]. This parameter is known and given by $\theta_{\max} = 2R_c/(cT_s)$, where $R_c$ is the cell radius and $T_s = 1/(N\Delta f)$ the sampling period. A simple way to counteract the effects of timing errors relies on the use of sufficiently long CPs comprising $N_G \geq \theta_{\max} + L$ sampling intervals. This leads to a *quasi-synchronous* network in which timing errors do not produce any IBI and only appear as phase shifts at the output of the receive discrete Fourier transform (DFT) unit. Although such a solution is normally adopted during IR, the CP of data symbols should be made just greater than the channel length to minimize unnecessary overhead. It follows that accurate timing estimates must be obtained during the ranging period in order to avoid IBI over the data section of the frame.

## B. Signal model

We denote by $Y_m(i_{q,\nu})$ the DFT output over the $i_{q,\nu}$th subcarrier of the $m$th OFDM symbol. Assuming for simplicity that the DSSs are perfectly aligned to the BS references, their signals do not contribute to $Y_m(i_{q,\nu})$. In contrast, the presence of uncompensated CFOs destroys orthogonality among ranging signals and gives rise to ICI. On the other hand, recalling that the CFOs are confined within a small fraction of the subcarrier spacing, the demodulated signals incur negligible phase rotations over one OFDM symbol and the resulting ICI can reasonably be neglected. Under the above assumptions, we may write

$$Y_m(i_{q,\nu}) \approx \sum_{k=1}^{K} c_k(m) e^{j2\pi m \varepsilon_k N_T/N} S_k(\theta_k, i_{q,\nu}) + n_m(i_{q,\nu}) \tag{2}$$

where $N_T = N + N_G$ denotes the duration of the cyclically extended OFDM symbols and $n_m(i_{q,\nu})$ accounts for Gaussian noise with zero mean and variance $\sigma^2$. In addition, we have defined

$$S_k(\theta_k, i_{q,\nu}) = e^{-j2\pi\theta_k i_{q,\nu}/N} H_k(i_{q,\nu}) \tag{3}$$

where

$$H_k(i_{q,\nu}) = \sum_{\ell=0}^{L-1} h_k(\ell) e^{-j2\pi\ell i_{q,\nu}/N} \tag{4}$$

is the $k$th channel frequency response over the $i_{q,\nu}$th subcarrier. The power received from the $k$th RSS over the ranging subcarriers is defined as

$$P_k = \frac{1}{QV} \sum_{q=0}^{Q-1} \sum_{\nu=0}^{V-1} |S_k(\theta_k, i_{q,\nu})|^2. \tag{5}$$

In the following sections we show how the quantities $\{Y_m(i_{q,\nu})\}$ can be exploited to identify the active codes and estimate the corresponding CFOs, timing errors and power levels. As anticipated, in doing so we adopt a multistage procedure in which frequency estimates are preliminarily obtained and are next used for timing and power estimation purposes. For the time being, we let $K \leq M-1$ and assume that the active RSSs use different codes over the considered subchannel [6], [8]. A method to identify possible collisions between RSSs employing the same code is presented in Sect. V.

## III. CODE DETECTION AND CFO ESTIMATION

We define an $M-$dimensional vector $\mathbf{Y}(i_{q,\nu}) = [Y_0(i_{q,\nu}), Y_1(i_{q,\nu}), \ldots, Y_{M-1}(i_{q,\nu})]^T$ which collects the $i_{q,\nu}$th DFT output across the ranging time-slot. Then, from (2) we have

$$\mathbf{Y}(i_{q,\nu}) = \sum_{k=1}^{K} S_k(\theta_k, i_{q,\nu})\mathbf{\Gamma}(\varepsilon_k)\mathbf{c}_k + \mathbf{n}(i_{q,\nu}) \tag{6}$$

where

$$\mathbf{\Gamma}(\varepsilon_k) = \text{diag}\{e^{j2\pi m \varepsilon_k N_T/N} ; m = 0, 1, \ldots, M-1\} \tag{7}$$

while $\mathbf{n}(i_{q,\nu}) = [n_0(i_{q,\nu}), n_1(i_{q,\nu}), \ldots, n_{M-1}(i_{q,\nu})]^T$ is Gaussian distributed with zero-mean and covariance matrix $\sigma^2\mathbf{I}_M$. Letting $\boldsymbol{\varepsilon} = [\varepsilon_1, \varepsilon_2, \ldots, \varepsilon_K]^T$ and $\boldsymbol{\theta} = [\theta_1, \theta_2, \ldots, \theta_K]^T$, we may rewrite $\mathbf{Y}(i_{q,\nu})$ in a more compact form as

$$\mathbf{Y}(i_{q,\nu}) = \mathbf{C}(\boldsymbol{\varepsilon})\mathbf{S}(\boldsymbol{\theta}, i_{q,\nu}) + \mathbf{n}(i_{q,\nu}) \tag{8}$$

where $\mathbf{C}(\boldsymbol{\varepsilon})$ is the following $M \times K$ matrix

$$\mathbf{C}(\boldsymbol{\varepsilon}) = [\mathbf{\Gamma}(\varepsilon_1)\mathbf{c}_1 \quad \mathbf{\Gamma}(\varepsilon_2)\mathbf{c}_2 \quad \cdots \quad \mathbf{\Gamma}(\varepsilon_K)\mathbf{c}_K] \tag{9}$$

and $\mathbf{S}(\boldsymbol{\theta}, i_{q,\nu}) = [S_1(\theta_1, i_{q,\nu}), S_2(\theta_2, i_{q,\nu}), \ldots, S_K(\theta_K, i_{q,\nu})]^T$. From (6) we see that $\mathbf{Y}(i_{q,\nu})$ is a superposition of frequency-rotated codes $\{\mathbf{\Gamma}(\varepsilon_k)\mathbf{c}_k\}_{k=1}^{K}$ embedded in white Gaussian noise and with random amplitudes $\{S_k(\theta_k, i_{q,\nu})\}_{k=1}^{K}$. This model has the same structure as measurements of multiple uncorrelated sources from an array of sensors. We can thus identify the active codes and their corresponding CFOs by applying subspace-based methods.

### A. Determination of the number of active codes

The first problem is to determine the number of received codes over the considered ranging subchannel. This task can be accomplished through the eigenvalue decomposition (EVD) of the correlation matrix $\mathbf{R}_Y = \text{E}\{\mathbf{Y}(i_{q,\nu})\mathbf{Y}^H(i_{q,\nu})\}$. In practice, however, $\mathbf{R}_Y$ is not available at the receiver. One common approach is to use the sample correlation matrix

$$\hat{\mathbf{R}}_Y = \frac{1}{QV}\sum_{\nu=0}^{V-1}\sum_{q=0}^{Q-1}\mathbf{Y}(i_{q,\nu})\mathbf{Y}^H(i_{q,\nu}) \tag{10}$$

 

which provides an unbiased and consistent estimate of $\mathbf{R}_Y$. Performing the EVD on $\hat{\mathbf{R}}_Y$ and arranging the corresponding eigenvalues $\hat{\lambda}_1 \geq \hat{\lambda}_2 \geq \cdots \geq \hat{\lambda}_M$ in non-increasing order, we can find an estimate $\hat{K}$ of the number of active codes through information-theoretic criteria. Two prominent solutions in this sense are based on the Akaike and MDL criteria. Here, we adopt the MDL approach which looks for the minimum of the following objective function [9]

$$\mathcal{F}(\tilde{K}) = \frac{1}{2}\tilde{K}(2M - \tilde{K})\ln(QV) - QV(M - \tilde{K})\ln\rho(\tilde{K}) \tag{11}$$

with

$$\rho(\tilde{K}) = \frac{\left(\prod_{i=\tilde{K}+1}^{M} \hat{\lambda}_i\right)^{\frac{1}{M-\tilde{K}}}}{\frac{1}{M-\tilde{K}}\sum_{i=\tilde{K}+1}^{M} \hat{\lambda}_i}. \tag{12}$$

Extensive numerical simulations indicate that a better estimate of $K$ is obtained by replacing the smallest eigenvalue $\hat{\lambda}_M$ with an estimate $\hat{\sigma}^2$ of the noise power. The latter can be obtained in various ways. One possible approach is based on the use of null subcarriers placed at the spectrum edges and reads

$$\hat{\sigma}^2 = \frac{1}{2MN_0}\sum_{m=0}^{M-1}\sum_{n=0}^{N_0-1}\left[|Y_m(n)|^2 + |Y_m(N-n-1)|^2\right] \tag{13}$$

where $Y_m(n)$ is the DFT output corresponding to the $n$th subcarrier of the $m$th OFDMA symbol.

### B. CFO estimation and code detection

Inspection of (6) reveals that the observation space can be decomposed into a signal subspace $\mathcal{S}_s$ spanned by the rotated codes $\{\mathbf{\Gamma}(\varepsilon_k)\mathbf{c}_k\}_{k=1}^{K}$ plus a noise subspace $\mathcal{S}_n$. Since $\mathcal{S}_n$ is the orthogonal complement of $\mathcal{S}_s$, each vector in $\mathcal{S}_s$ is orthogonal to any other vector in $\mathcal{S}_n$. For simplicity, we assume that the number of active codes has been correctly estimated and denote by $\{\hat{\mathbf{u}}_1, \hat{\mathbf{u}}_2, \ldots, \hat{\mathbf{u}}_M\}$ the eigenvectors of $\hat{\mathbf{R}}_Y$ corresponding to the ordered eigenvalues $\hat{\lambda}_1 \geq \hat{\lambda}_2 \geq \cdots \geq \hat{\lambda}_M$. The MUSIC algorithm relies on the fact that the eigenvectors $\{\hat{\mathbf{u}}_{K+1}, \hat{\mathbf{u}}_{K+2}, \ldots, \hat{\mathbf{u}}_M\}$ associated with the $M - K$ smallest eigenvalues of $\hat{\mathbf{R}}_Y$ form an *estimated* basis of $\mathcal{S}_n$ and, accordingly, they are *approximately* orthogonal to all vectors in the signal space [10]. An estimate of $\varepsilon_k$ is thus obtained by minimizing the projection of $\mathbf{\Gamma}(\tilde{\varepsilon})\mathbf{c}_k$ onto the subspace spanned by the columns of $\hat{\mathbf{U}}_n = [\hat{\mathbf{u}}_{K+1}\ \hat{\mathbf{u}}_{K+2}\ \cdots\ \hat{\mathbf{u}}_M]$. This leads to the following estimation algorithm

$$\hat{\varepsilon}_k = \arg\max_{\tilde{\varepsilon}} \{\Psi_k(\tilde{\varepsilon})\} \tag{14}$$

with

$$\Psi_k(\tilde{\varepsilon}) = \frac{1}{\left\| \hat{\mathbf{U}}_n^H \mathbf{\Gamma}(\tilde{\varepsilon})\mathbf{c}_k \right\|^2}. \tag{15}$$

In principle, CFO recovery must only be accomplished for the active codes. However, since at this stage the BS has no knowledge as to which codes are being employed over the considered subchannel, frequency estimates $\{\hat{\varepsilon}_1, \hat{\varepsilon}_2, \ldots, \hat{\varepsilon}_M\}$ must be evaluated for the complete set $\mathcal{C}$. Next, the problem arises of how to identify the codes that are actually active. The proposed identification algorithm looks for the $K$ largest values in the set $\{\Psi_\ell(\hat{\varepsilon}_\ell)\}_{\ell=1}^M$, say $\{\Psi_{u_k}(\hat{\varepsilon}_{u_k})\}_{k=1}^K$, and declares as *active* the corresponding codes $\{\mathbf{c}_{u_k}\}_{k=1}^K$. The CFO estimates are eventually found as $\hat{\boldsymbol{\varepsilon}}_u = [\hat{\varepsilon}_{u_1}, \hat{\varepsilon}_{u_2}, \ldots, \hat{\varepsilon}_{u_K}]^T$. In the sequel, we refer to (14) as the MUSIC-based frequency estimator (MFE), while the described identification algorithm is called the MUSIC-based code detector (MCD).

## C. Remarks

1) A fundamental assumption behind the MUSIC estimator is that the dimension of $\mathcal{S}_n$ is at least unitary. This implies $K < M$, which explains why the number of RSSs over the same ranging subchannel should not exceed $M-1$.

2) Let $K \leq M-1$ and assume that two or more RSSs share the same ranging code over the considered subchannel. In such a case, although the MDL can still provide the exact number of active RSSs, the MCD is not capable of identifying the corresponding codes as it implicitly assumes that any given code is used by no more than one RSS. As an example, let $M=4$ and $K=3$ with one RSS employing code $\mathbf{c}_1$ and the other two RSSs $\mathbf{c}_2$. In this situation, it is likely that the MCD declares as active $\mathbf{c}_1$ and $\mathbf{c}_2$ plus a third code ($\mathbf{c}_3$ or $\mathbf{c}_4$) which is actually turned off and corresponds to a *phantom RSS*. In a similar way, the MFE will provide three CFO estimates, two of which are associated with active users while the remaining one corresponds to the phantom RSS. It is worth observing that, when $\hat{K} = K$, the presence of a phantom RSS always implies the mis-detection of an active user, which is referred to as *undetected RSS*.

3) In multiple frequency estimation problems the ESPRIT [12] represents a valid alternative to the MUSIC [13]. The main advantage of ESPRIT is that it provides estimates of the signal parameters in closed-form without requiring any time consuming grid-search. A basic assumption behind this technique is the rotational invariance property of the observation vectors, which is guaranteed in the presence of complex exponentials in noise. Unfortunately, in general, the rotated codes $\{\mathbf{\Gamma}(\varepsilon_k)\mathbf{c}_k\}_{k=1}^K$ in general do not satisfy the invariance property. In this case, the ESPRIT cannot be used and we must apply the MUSIC or alternative reduced-complexity schemes like root-MUSIC or the minimum-norm method [14].

4) A key issue is the maximum CFO that the MFE can handle. To simplify the analysis, assume that the ranging codes belong to the Fourier basis of order $M$, i.e.,

$$c_k(m) = e^{j2\pi m(k-1)/M}, \qquad 0 \le m \le M-1 \tag{16}$$

for $k = 1, 2, \ldots, M$. Then, in Appendix it is shown that identifiability of the rotated codes $\{\mathbf{\Gamma}(\varepsilon_k)\mathbf{c}_k\}_{k=1}^K$ is guaranteed as long as the normalized CFOs are smaller than $N/(2MN_T)$ in magnitude. On the other hand, assume that $\varepsilon_k = N/(2MN_T)$ and $\varepsilon_{k+1} = -N/(2MN_T)$. In such a case, the received codes $\mathbf{\Gamma}(\varepsilon_k)\mathbf{c}_k$ and $\mathbf{\Gamma}(\varepsilon_{k+1})\mathbf{c}_{k+1}$ are identical and there is no way to distinguish among them. The above facts say that the acquisition range of MFE is $|\varepsilon_k| < N/(2MN_T)$. Although this result cannot be easily extended to other code designs, extensive simulations indicate that it also applies to Walsh-Hadamard binary codes.

5) The accuracy of MFE can be assessed with the methods developed in [15]. Specifically, at high SNR values and for large data records (i.e., large values of $QV$), the CFO estimation errors are nearly Gaussian distributed with zero mean and variances

$$\mathrm{E}\{(\hat{\varepsilon}_k - \varepsilon_k)^2\} = \frac{\sigma^2 N^2}{8\pi^2 QV N_T^2 P_k} \cdot \frac{1}{\mathbf{d}_k^H(\varepsilon_k)\mathbf{C}^\perp(\boldsymbol{\varepsilon})\mathbf{d}_k(\varepsilon_k)} \tag{17}$$

where $\mathbf{d}_k(\varepsilon_k)$ is an $M-$dimensional vector with entries $\{mc_k(m)e^{j2\pi m\varepsilon_k N_T/N}\}_{m=0}^{M-1}$ and $\mathbf{C}^\perp(\boldsymbol{\varepsilon}) = \mathbf{I}_M - \mathbf{C}(\boldsymbol{\varepsilon})[\mathbf{C}^H(\boldsymbol{\varepsilon})\mathbf{C}(\boldsymbol{\varepsilon})]^{-1}\mathbf{C}^H(\boldsymbol{\varepsilon})$.

6) It is interesting to assess the computational requirement of MFE and MCD. Computing $\hat{\mathbf{R}}_Y$ in (10) approximately involves $2QVM^2$ real multiplications plus the same number of real additions. In writing these figures we have borne in mind that $\hat{\mathbf{R}}_Y$ is Hermitian and, accordingly, only the entries $[\hat{\mathbf{R}}_Y]_{m_1,m_2}$ with $m_1 \geq m_2$ need be computed. Performing the EVD on a *real-valued* $M \times M$ matrix needs $12M^3$ floating operations (flops) [16]. For the *complex-valued* matrix $\hat{\mathbf{R}}_Y$, we approximate the flop count as $72M^3$ by pessimistically treating every operation as it were a complex multiplication. Finally, evaluating the metric $\Psi_k(\tilde{\varepsilon})$ in (15) requires $(4M + 2)(M - K)$ real products plus $4M(M - K)$ real additions for each $\tilde{\varepsilon}$. Denoting by $N_\varepsilon$ the number of candidate values $\tilde{\varepsilon}$ and observing that the maximization in (14) must be performed over the entire code set $\{\mathbf{c}_1, \mathbf{c}_2, \ldots, \mathbf{c}_M\}$, it follows that a total of $2N_\varepsilon M(4M + 1)(M - K))$ operations are required to find $\hat{\varepsilon}_u$. The overall number of flops required by MFE and MCD in the considered ranging subchannel are summarized in the first row of Tab. I.

## IV. Estimation of the timing delays and power levels

After code detection and CFO recovery, the BS must acquire information about the timing delays and power levels of all ranging signals. This problem is addressed by resorting to a two-step procedure in which estimates of $\mathbf{S}(\boldsymbol{\theta}, i_{q,\nu})$ are firstly computed over all ranging subcarriers and are subsequently exploited to obtain the timing errors and power levels. To simplify the notation, in the following derivations the indices $\{u_k\}_{k=1}^{\hat{K}}$ of the detected codes are relabeled according to the map $u_k \rightarrow k$.

### A. Timing recovery

Let $\hat{K} = K$ and assume that the accuracy of the CFO estimates is such that $\hat{\boldsymbol{\varepsilon}} \simeq \boldsymbol{\varepsilon}$. Then, the maximum likelihood (ML) estimate of $\mathbf{S}(\boldsymbol{\theta}, i_{q,\nu})$ based on the model (8) is found to be

$$\hat{\mathbf{S}}(i_{q,\nu}) = [\hat{\mathbf{C}}^H(\hat{\boldsymbol{\varepsilon}})\hat{\mathbf{C}}(\hat{\boldsymbol{\varepsilon}})]^{-1}\hat{\mathbf{C}}^H(\hat{\boldsymbol{\varepsilon}})\mathbf{Y}(i_{q,\nu}) \tag{18}$$

where $\hat{\mathbf{C}}(\hat{\boldsymbol{\varepsilon}}) = [\boldsymbol{\Gamma}(\hat{\varepsilon}_1)\mathbf{c}_1 \;\; \boldsymbol{\Gamma}(\hat{\varepsilon}_2)\mathbf{c}_2 \; \cdots \; \boldsymbol{\Gamma}(\hat{\varepsilon}_{\hat{K}})\mathbf{c}_{\hat{K}}]$ and we have omitted the functional dependence of $\mathbf{S}(\boldsymbol{\theta}, i_{q,\nu})$ on $\boldsymbol{\theta}$. Substituting (8) into (18) yields

$$\hat{\mathbf{S}}(i_{q,\nu}) = \mathbf{S}(i_{q,\nu}) + \boldsymbol{\psi}(i_{q,\nu}) \tag{19}$$

where $\boldsymbol{\psi}(i_{q,\nu}) = [\psi_1(i_{q,\nu}), \psi_2(i_{q,\nu}), \ldots, \psi_K(i_{q,\nu})]^T$ is a zero-mean disturbance vector with co-variance matrix $\sigma^2[\hat{\mathbf{C}}^H(\hat{\boldsymbol{\varepsilon}})\hat{\mathbf{C}}(\hat{\boldsymbol{\varepsilon}})]^{-1}$. From (19) and (3) it follows that the entries of $\hat{\mathbf{S}}(i_{q,\nu})$ can be modeled as

$$\hat{S}_k(i_{q,\nu}) = e^{-j2\pi\theta_k i_{q,\nu}/N} H_k(i_{q,\nu}) + \psi_k(i_{q,\nu}) \qquad 1 \le k \le K \tag{20}$$

for any $q \in \{0, 1, \dots, Q-1\}$ and $\nu \in \{0, 1, \dots, V-1\}$. The above equation indicates that, under the assumptions of ideal CFO recovery and code detection, $\hat{S}_k(i_{q,\nu})$ is only contributed by the $k$th RSS. Hence, we can use the quantities $\{\hat{S}_k(i_{q,\nu})\}$ to get timing estimates individually for each RSS.

To see how this comes about, we recall that in practical multicarrier systems the channel gains over contiguous subcarriers are highly correlated and their values are almost the same. Hence, letting $H_k(i_{q,\nu}-1) \simeq H_k(i_{q,\nu})$ and neglecting for simplicity the noise contribution, from (20) we have

$$\hat{S}_k(i_{q,\nu}-1)\hat{S}_k^*(i_{q,\nu}) \simeq |H_k(i_{q,\nu})|^2 \, e^{j2\pi\theta_k/N} \; . \tag{21}$$

The above result indicates that a timing estimate can be obtained in the form

$$\hat{\theta}_k = \text{round} \left\{ \frac{N}{2\pi} \arg \left[ \sum_{q=0}^{Q-1} \sum_{\nu=1}^{V-1} \hat{S}_k(i_{q,\nu}-1)\hat{S}_k^*(i_{q,\nu}) \right] \right\} \tag{22}$$

where the round operation is justified by the fact that $\theta_k$ is integer-valued.

As discussed in [11], IBI is present during the data section of the frame whenever the timing error $\hat{\theta}_k - \theta_k$ lies outside the interval $J_{\Delta\theta} = [L - N_{GD} - 1, 0]$, where $N_{G,D}$ is the CP length of the OFDMA symbols during the data section of the frame. Intuitively speaking, the probability of occurrence of IBI can be reduced by shifting the expected value of the timing error toward the middle point of $J_{\Delta\theta}$, which is given by $(L - N_{GD} - 1)/2$. This leads to a refined timing estimate $\hat{\theta}_k^{(f)} = \hat{\theta}_k - \mu_k + (L - N_{GD} - 1)/2$, with $\mu_k = \text{E}\{\hat{\theta}_k - \theta_k\}$. Extensive simulations indicate that $\mu_k$ equals the mean delay associated to $\mathbf{h}_k$. This parameter is generally unknown at the receiver, but can roughly be approximated by $(L-1)/2$. Combining the above results with (22) produces

$$\hat{\theta}_k^{(f)} = \text{round} \left\{ \frac{N}{2\pi} \arg \left[ \sum_{q=0}^{Q-1} \sum_{\nu=1}^{V-1} \hat{S}_k(i_{q,\nu}-1)\hat{S}_k^*(i_{q,\nu}) \right] - \frac{N_{GD}}{2} \right\} \tag{23}$$

which is referred to as the ad-hoc timing estimator (AHTE).

An alternative approach to get timing estimates from the quantities $\{\hat{S}_k(i_{q,\nu})\}$ is illustrated in [17] by resorting to the least-squares criterion. Compared to this solution, AHTE is simpler to implement and can handle a maximum timing offset of $N/2$, whereas in [17] the estimation range depends on $Q$ and $L$ and is typically much smaller than $N/2$.

## B. Power level estimation

Collecting (3) and (20), we observe that $\hat{S}_k(i_{q,\nu})$ is an unbiased estimate of $S_k(i_{q,\nu})$ with variance

$$\sigma_k^2 = \sigma^2 \cdot [\hat{\mathbf{T}}^{-1}(\hat{\varepsilon})]_{k,k} \tag{24}$$

where $\hat{\mathbf{T}}(\hat{\varepsilon}) = \hat{\mathbf{C}}^H(\hat{\varepsilon})\hat{\mathbf{C}}(\hat{\varepsilon})$. An unbiased estimate of $|S_k(i_{q,\nu})|^2$ is thus given by $|\hat{S}_k(i_{q,\nu})|^2 - \sigma_k^2$. This fact, together with (5) and (24), leads to the following ad-hoc power estimator (AHPE)

$$\hat{P}_k = \frac{1}{QV} \sum_{q=0}^{Q-1} \sum_{\nu=0}^{V-1} |\hat{S}_k(i_{q,\nu})|^2 - \hat{\sigma}^2 \cdot [\hat{\mathbf{T}}^{-1}(\hat{\varepsilon})]_{k,k} \tag{25}$$

where $\hat{\sigma}^2$ is an estimate of $\sigma^2$.

In case of ideal frequency and noise power estimation (i.e., $\hat{\varepsilon} = \varepsilon$ and $\hat{\sigma}^2 = \sigma^2$), it can be shown that AHPE provides unbiased estimates with variance

$$\text{var}\{\hat{P}_k\} = \frac{\sigma_k^2(2P_k + \sigma_k^2)}{QV} \quad \text{for} \ \ k = 1, 2, \dots, K. \tag{26}$$

## C. Complexity issues

In assessing the computational load of AHTE and AHPE, it is useful to distinguish between a first stage, leading to vectors $\hat{\mathbf{S}}(i_{q,\nu})$ in (18), and a second stage where these vectors are used to compute the timing and power estimates. We begin by observing that $\hat{\mathbf{S}}(i_{q,\nu})$ is the solution of a linear system $\hat{\mathbf{C}}^H(\hat{\varepsilon})\hat{\mathbf{C}}(\hat{\varepsilon})\hat{\mathbf{S}}(i_{q,\nu}) = \hat{\mathbf{C}}^H(\hat{\varepsilon})\mathbf{Y}(i_{q,\nu})$. Assuming that the entries of $\hat{\mathbf{C}}(\hat{\varepsilon})$ are available, computing the Hermitian matrix $\hat{\mathbf{C}}^H(\hat{\varepsilon})\hat{\mathbf{C}}(\hat{\varepsilon})$ involves $4K^2M$ flops while evaluating $\hat{\mathbf{C}}^H(\hat{\varepsilon})\mathbf{Y}(i_{q,\nu})$ needs $8KM$ flops for any pair $(q, \nu)$. Each linear system is efficiently solved with approximately $2K^2(K+6)$ flops by exploiting the Cholesky factorization of $\hat{\mathbf{C}}^H(\hat{\varepsilon})\hat{\mathbf{C}}(\hat{\varepsilon})$ [16]. A total of $4K^2M + 2KQV(K^2 + 6K + 4M)$ flops are thus required to obtain $\hat{\mathbf{S}}(i_{q,\nu})$ for $q \in \{0, 1, \dots, Q-1\}$ and $\nu \in \{0, 1, \dots, V-1\}$. In the second stage of the estimation process, timing and power estimates are obtained through (23) and (25), respectively. Computing $\hat{\theta}_k$ from (23) involves $8Q(V-1)$ flops for any active RSS, while $4QV$ additional flops are needed to evaluate $\hat{P}_k$ in (25). The overall requirement of AHTE and APHE in the considered ranging subchannel is summarized in the second row of Tab. I. In writing these figures we have not considered the noise power estimate (13), which is computed off-line with negligible complexity.

## V. Collision detection

So far, we have neglected possible collisions between RSSs that choose the same ranging opportunity. This implies that each subchannel is accessed by no more than $M - 1$ RSSs which employ different codes. Although a proper system design can reduce the risk of a collision, such event occurs with some non-zero probability. As mentioned previously, if $K \geq M$ the MFE cannot work properly since in this case the noise subspace reduces to the null vector. Another critical situation is represented by the presence of pairs of phantom and undetected RSSs, which is likely to occur when $K \leq M - 1$ and the same code is shared by two or more RSSs. If the undetected RSS employs the same code of a detected RSS, the former will adjust its transmission parameters according to the response message transmitted to the latter. Such adjustment may have detrimental effects as it is based on incorrect synch information. On the other hand, the presence of a phantom RSS is not a big problem as the code associated to the corresponding response message will not recognized by any of the active RSSs. The above discussion indicates that, although it is not strictly necessary to notify the collision status to collided RSSs, collision events must be detected to avoid that incorrect synch information be trasmitted to the active RSSs. For this purpose, we consider the following vectors

$$\Delta \mathbf{Y}(i_{q,\nu}) = \mathbf{Y}(i_{q,\nu}) - \hat{\mathbf{C}}(\hat{\boldsymbol{\varepsilon}})\hat{\mathbf{S}}(i_{q,\nu}) \tag{27}$$

for $q \in \{0, 1, \ldots, Q - 1\}$ and $\nu \in \{0, 1, \ldots, V - 1\}$. From (8), we conjecture that the entries of $\Delta \mathbf{Y}(i_{q,\nu})$ will have small amplitudes if $\hat{\mathbf{C}}(\hat{\boldsymbol{\varepsilon}})$ and $\hat{\mathbf{S}}(i_{q,\nu})$ are good approximations of $\mathbf{C}(\boldsymbol{\varepsilon})$ and $\mathbf{S}(i_{q,\nu})$, respectively. Such situation occurs when the estimates $\hat{K}$ and $\hat{\boldsymbol{\varepsilon}}$ are sufficiently accurate. Indeed, using standard computations it can be shown that

$$\mathrm{E}\{\|\Delta \mathbf{Y}(i_{q,\nu})\|^2\} = \sigma^2(M - \hat{K}) + \delta(\hat{\boldsymbol{\varepsilon}}, \boldsymbol{\varepsilon}, i_{q,\nu}) \tag{28}$$

where

$$\delta(\hat{\varepsilon}, \varepsilon, i_{q,\nu}) = \begin{cases} 0 & \text{if } \hat{K} = K \text{ and } \hat{\varepsilon} = \varepsilon \\ \left\| \hat{\mathbf{Z}}^H(\hat{\varepsilon}) \mathbf{C}(\varepsilon) \mathbf{S}(i_{q,\nu}) \right\|^2 & \text{otherwise} \end{cases} \quad (29)$$

and $\hat{\mathbf{Z}}(\hat{\varepsilon})\hat{\mathbf{Z}}^H(\hat{\varepsilon})$ is the Cholesky factorization of the $M \times M$ matrix $\mathbf{I}_M - \hat{\mathbf{C}}(\hat{\varepsilon})[\hat{\mathbf{C}}^H(\hat{\varepsilon})\hat{\mathbf{C}}(\hat{\varepsilon})]^{-1}\hat{\mathbf{C}}^H(\hat{\varepsilon})$. Inspection of (29) reveals that $\delta(\hat{\varepsilon}, \varepsilon, i_{q,\nu})$ is minimum when $K$ and $\varepsilon$ are perfectly estimated, while much larger values are expected in the presence of colliding RSSs due to the poor quality of $\hat{K}$ and $\hat{\varepsilon}$. This suggests the use of $\delta(\hat{\varepsilon}, \varepsilon, i_{q,\nu})$ as a collision detection metric. Unfortunately, this quantity is unknown and must be estimated in some manner. Replacing the expectation of $\|\Delta\mathbf{Y}(i_{q,\nu})\|^2$ in (28) with the corresponding ensemble average, an estimate of $\delta(\hat{\varepsilon}, \varepsilon, i_{q,\nu})$ is obtained as

$$\hat{\delta} = \frac{1}{QV} \sum_{q=0}^{Q-1} \sum_{\nu=0}^{V-1} \|\Delta\mathbf{Y}(i_{q,\nu})\|^2 - \hat{\sigma}^2(M - \hat{K}) \quad (30)$$

where $\hat{\sigma}^2$ is expressed in (13). Hence, a collision status is declared to occur whenever $\hat{\delta}$ exceeds a specified threshold $\eta$. In that case, the BS does not send any response message to the users on the considered subchannel as their estimated synch parameters and power levels are regarded as not sufficiently reliable. The RSSs that do not find their information repeat the ranging process in the next frame using a different ranging opportunity. If $\hat{\delta} < \eta$, the BS considers the users' parameters as accurately estimated and sends a response message for all detected codes in the considered subchannel. In the sequel, we refer to the above procedure as the ad-hoc collision detector (AHCD). Clearly, $\eta$ must be properly designed to achieve a reasonable trade-off between the probability of declaring a collision when in fact it is not present (false alarm) and the probability of not detecting a collision when in fact it is present (mis-detection). Since computing such probabilities by theoretical analysis appears a formidable task, numerical simulations can be used in practice for the design of $\eta$.

In assessing the computational load of AHCD, we let $\hat{K} = K$ and observe that $\hat{\mathbf{S}}(i_{q,\nu})$ is available at the BS as an output of AHTE. Hence, computing $\Delta\mathbf{Y}(i_{q,\nu})$ in (27) needs $8KM$ flops for any value of $q$ and $\nu$ while $4M - 1$ flops are required to get $\|\Delta\mathbf{Y}(i_{q,\nu})\|^2$. Finally, evaluating $\hat{\delta}$ in (30) involves $QV$ real additions. The resulting complexity for the considered subchannel is shown in the third line of Tab. I.

## VI. SIMULATION RESULTS

### A. System parameters

The investigated OFDMA system is based on the IEEE 802.16e standard for wireless MANs. The DFT size is $N = 1024$ and the sampling period is $T_s = 87.5 \ ns$, corresponding to a subcarrier distance of $1/(NT_s) = 11.16$ kHz. We assume that $N_0 = 80$ subcarriers are placed at both edges of the signal spectrum. The number of modulated subcarriers is thus $N_U = 864$ while the uplink bandwidth is approximately $B_w = N_U/(NT_s) = 9.7$ MHz. A total of $N_R = 144$ subcarriers are reserved for ranging. They are divided into $R = 18$ subchannels, each comprising $Q = 4$ tiles uniformly spaced over the signal spectrum at a distance of $N_U/Q = 216$. The number of subcarriers in any tile is $V = 2$. The remaining $N_U - N_R = 720$ subcarriers are grouped into 15 data subchannels, each composed by 48 subcarriers. A ranging time-slot includes $M = 4$ OFDMA symbols. Hence, the number of ranging opportunities in each frame is $N_{\text{total}} = R(M-1) = 54$. The ranging codes are taken from the Fourier basis of order 4, while DSS data symbols belong to a QPSK constellation.

The discrete-time CIRs have maximum order $L = 14$. Their entries are modeled as independent and circularly symmetric Gaussian random variables with zero-mean and an exponential power delay profile, i.e.,

$$\mathrm{E}\{|h_k(\ell)|^2\} = \sigma_{h,k}^2 \cdot \exp(-\ell/L_k), \qquad \ell = 0, 1, \ldots, L_k - 1 \tag{31}$$

where $\{L_k\}_{k=1}^K$ are taken from the set $\{8, 9, \ldots, 14\}$ with equal probability while $\sigma_{h,k}^2 = (1 - e^{-1})/(1 - e^{-1/L_k})$ so as to have $\mathrm{E}\{\|\mathbf{h}_k\|^2\} = 1$ for all active users. In this way, the powers $P_k$ of the received signals are random variables with unit mean. Channels of different users are statistically independent of each other. We consider a cell radius of 1.5 km, corresponding to a maximum propagation delay of $\theta_{\max} = 114$ sampling periods. Ranging symbols are preceded by a CP of length $N_G = 128$ in order to avoid IBI. The normalized CFOs are uniformly distributed over the interval $[-\varepsilon_{\max}, \varepsilon_{\max}]$ and vary at each run. Recalling that the estimation range of MFE is $|\varepsilon_k| < N/(2MN_T)$, throughout simulations we set $\varepsilon_{\max} \leq 0.1$ for the RSSs. The number of CFO trial values in (14) is $N_\varepsilon = 400$, corresponding to an MFE frequency resolution of $2\varepsilon_{\max}/N_\varepsilon \leq 5 \cdot 10^{-4}$. For the DSSs we fix $\varepsilon_{\max} = 0.02$, while the maximum timing error is limited to 48 samples.

Unless otherwise specified, we consider a static scenario where channel coefficients are generated at each simulation run and kept fixed over an entire time-slot. In evaluating the accuracy of the synchronization and code detection algorithms, we assume that all RSSs attempt their ranging process simultaneously at the first time-slot choosing different ranging opportunities. Collision events are simulated only to assess the performance of AHCD. Comparisons are made with the ranging scheme proposed by Fu, Li and Minn (FLM) in [8] under a common simulation set-up. This includes the same number of ranging subcarriers, ranging subchannels and data subchannels, as well as the same transmitted energy from each user terminal. However, as FLM can support $M$ different RSS over the same subchannel, the total number of ranging opportunities is 72 with FLM and 54 with our scheme.

In all subsequent simulations, the same number $K$ of RSSs is present in *each* ranging subchannels. This implies that a total of $\gamma_R = KR$ RSSs are simultaneously active in the system. Note that letting $K = 3$ in our ranging scheme corresponds to a fully-loaded system where all ranging opportunities are employed. We fix the number of DSSs to $\gamma_D = 10$, although extensive simulations indicate that the system performance is only marginally affected by this parameter.

### B. Performance evaluation

*1) Multiuser code detection:* We begin by investigating the performance of MCD in terms of mis-detection and false alarm probabilities, say $P_{md}$ and $P_{fa}$. Fig. 1 illustrates $P_{md}$ as a function of $\text{SNR} = 1/\sigma^2$. The number of active RSSs in each subchannel is either 2 or 3 while the maximum normalized CFO is 0.05. The FLM code detector declares the $k$th ranging code as active provided that the quantity

$$\mathcal{Z}_k = \frac{1}{QVM^2} \sum_{q=0}^{Q-1} \sum_{\nu=0}^{V-1} \left| \mathbf{c}_k^H \mathbf{Y}(i_{q,\nu}) \right|^2 \tag{32}$$

exceeds a suitable threshold $\eta_{FLM}$ which is proportional to the estimated noise power $\hat{\sigma}^2$. The results of Fig. 1 indicate that MCD performs remarkably better than FLM. As expected, the system performance deteriorates as $K$ approaches $M$. The reason is that increasing $K$ reduces the noise subspace dimension, thereby degrading the accuracy of MCD.

Fig. 2 shows $P_{fa}$ versus SNR for $K = 1$ or 2 and $\varepsilon_{\max} = 0.05$. As is seen, MCD outperforms FLM at SNR values greater than of 6 dB. Interestingly, when using FLM the probability of false alarm increases with SNR. Such behavior can be explained by observing that the threshold $\eta_{FLM}$ is proportional to $\hat{\sigma}^2$, so that it becomes smaller and smaller as the SNR increases.

*2) Frequency estimation:* Fig. 3 illustrates the root mean-square-error (RMSE) of the frequency estimates obtained with MFE vs. SNR for $K = 2$ or 3 and $\varepsilon_{\max} = 0.05$. The theoretical analysis in (17) is also shown for comparison. As it is seen, the frequency RMSE is approximately 2 dB worse than its predicted value. The reason is that the result (17) is accurate only for large values of $QV$ (large data records), while in our simulation set-up $QV$ is limited to 8. Again, the system performance deteriorates as $K$ increases. Nevertheless, the accuracy of MFE is satisfactory at all SNR values of practical interest. Indeed, an RMSE of $10^{-2}$ is obtained even with $K = 3$ if SNR $> 13$ dB.

The impact of the CFOs on the performance of MFE is assessed in Fig. 4, where the frequency RMSE is shown as a function of $\varepsilon_{\max}$. The SNR is fixed to 16 dB while $K$ is still 2 or 3. We see that the estimation accuracy is only marginally affected by $\varepsilon_{\max}$.

*3) Timing recovery:* The performance of the timing estimators is assessed by measuring the probability $P(\epsilon)$ of a timing error event. The latter occurs whenever the estimate $\hat{\theta}_k^{(f)}$ gives rise to IBI during the data section of the frame. This amounts to saying that the timing error $\Delta\hat{\theta}_k^{(f)} = \hat{\theta}_k^{(f)} - \theta_k$ is larger than zero or smaller than $L - N_{GD} - 1$, where $N_{GD} = 48$ is the CP length during the data transmission period. Note that the mean-shift $N_{GD}/2$ employed in (23) is also applied to the FLM timing estimator in order to reduce $P(\epsilon)$. Fig. 5 illustrates $P(\epsilon)$ vs. SNR as obtained with AHTE and FLM. The number of active codes in each ranging subchannel is $K = 2$ or 3 while $\varepsilon_{\max} = 0.05$. At practical SNR values, we see that AHTE provides much better results than FLM.

In Fig. 6 the timing estimators are compared in terms of their sensitivity to CFOs. For this purpose, $P(\epsilon)$ is shown as a function of $\varepsilon_{\max}$ for $K = 2$ or 3 and SNR $= 16$ dB. Again, we see that AHTE outperforms FLM, even though the latter is more robust to CFOs.

*4) Power estimation:* Fig. 7 illustrates the RMSE of the power estimates as obtained with AHPE and the FLM power estimator. In the latter case, the quantity $\hat{P}_k$ is computed as $\hat{P}_k = \mathcal{Z}_k/(QV) - \hat{\sigma}^2/M$ with $\mathcal{Z}_k$ as given in (32). The number of active RSSs in each subchannel is either 2 or 3 and $\varepsilon_{\max} = 0.05$. The theoretical analysis (26) is also shown for comparison.

Good agreement between simulation and theory is obtained for $K = 2$, while a loss of 2 dB is observed with $K = 3$. The reason is that (26) has been derived assuming perfect knowledge of the frequency offsets, while in practice the accuracy of the CFO estimates degrades with $K$. At low SNR values, both AHPE and FLM provide similar results, but the former takes the lead as the SNR grows large.

*5) Impact of channel variations:* All previous measurements have been conducted by keeping the channel responses fixed during one simulation run. In order to assess the impact of channel variations on the system performance, we now consider time-varying channel taps generated by passing white Gaussian noise through a third-order low-pass Butterworth filter. The 3-dB bandwidth of the filter is taken as a measure of the Doppler bandwidth $B_D = v f_c / c$, where $v$ denotes the mobile speed, $f_c = 2.5$ GHz is the carrier frequency and $c$ the speed of light. Figs. 8 and 9 illustrate the accuracy of MFE and AHTE, respectively, as a function of $v$ with SNR = 16 dB. Again, the number of active codes in each ranging subchannel is $K = 2$ or 3 while $\varepsilon_{\max} = 0.05$. The proposed scheme exhibits a remarkable robustness against channel variations and can handle a mobile speed of 30 m/s with negligible loss with respect to a static scenario.

*6) Collision detection:* Fig. 10 illustrates the performance of AHCD is terms of false alarm ($P_{fa}$) and mis-detection ($P_{md}$) probabilities as a function of the threshold $\eta$. The false alarm is measured in a scenario where two RSSs employing different ranging opportunities are active in each subchannel. Vice versa, $P_{md}$ is obtained with $K = 3$ and assuming that two RSSs in each subchannel choose the same code. In both situations the SNR is fixed to 16 dB while $\varepsilon_{\max} = 0.05$. As mentioned in Sect. V, it is important that the probability of mis-detecting a collision event is kept as low as possible in order to avoid that RSSs sharing the same ranging opportunity adjust their transmission parameters on the basis of incorrect synch information. The results shown in Fig. 10 indicate that a $P_{md}$ and $P_{fa}$ of $2 \cdot 10^{-3}$ can be obtained by setting $\eta = 0.05$.

### C. Computational complexity

It is interesting to compare the proposed ranging scheme with the FLM approach in terms of processing requirement in the considered simulation scenario. Assuming that $K = 2$ RSSs are active (on average) in each ranging subchannel, from Table I it follows that 114,000 flops are approximately needed by MCD and MFE for each ranging subchannel, while 1,280 flops are required by AHTE and AHPE. The complexity evaluation for FLM is performed in [8], where it is shown that code detection involves 1,088 flops while more than 3 Mflops are necessary for timing recovery in each subchannel. These results indicate that our scheme allows a significant complexity saving with respect to FLM, with a reduction of the overall number of flops by a factor 26.

## VII. Conclusions

We have presented a new ranging method for OFDMA systems where uplink signals arriving at the base station are plagued by frequency errors in addition to timing misalignments. The synchronization parameters of all ranging users are estimated in a decoupled fashion with affordable complexity. This is accomplished through a multistage procedure, where user identification and CFO estimation is performed first by means of a subspace decomposition approach. Frequency estimates are next employed for timing recovery and power level estimation. A simple method to detect possible collisions between RSSs employing the same ranging opportunity is also investigated. Compared to previous techniques, the proposed scheme exhibits improved accuracy with reduced complexity. Computer simulations indicate that the system performance is satisfactory even in the presence of frequency errors as large as 10% of the subcarrier spacing. The proposed approach can be used to enhance the ranging process of commercial IEEE 802.16-based OFDMA systems.

## Appendix

This Appendix establishes necessary conditions for code identifiability by means of MCD. To make the problem analytically tractable, the ranging codes are taken from the Fourier basis of order $M$ as expressed in (16). From (7) and (16) we see that matrix $\mathbf{C}(\boldsymbol{\varepsilon})$ in (9) has the following Vandermonde structure

$$\mathbf{C}(\boldsymbol{\varepsilon}) = \begin{bmatrix} 1 & 1 & \cdots & 1 \\ z_1 & z_2 & \cdots & z_K \\ \vdots & \vdots & \ddots & \vdots \\ z_1^{M-1} & z_2^{M-1} & \cdots & z_K^{M-1} \end{bmatrix} \tag{33}$$

with

$$z_k = e^{j2\pi[(k-1)/M + \varepsilon_k N_T/N]} \quad \text{for} \quad k = 1, 2, \dots, K. \tag{34}$$

Recalling that any given code can be univocally identified as long as it is linearly independent of all other codes, it follows that $\mathbf{C}(\varepsilon)$ must be full-rank. On the other hand, for a Vandermonde matrix the full-rank condition is met if and only if $z_{k_1} \neq z_{k_2}$ for $k_1 \neq k_2$. Bearing in mind (34), it is easily seen that $z_{k_1} \neq z_{k_2}$ is equivalent to putting

$$\frac{k_1}{M} + \frac{\varepsilon_{k_1} N_T}{N} \neq \ell + \frac{k_2}{M} + \frac{\varepsilon_{k_2} N_T}{N} \tag{35}$$

for any integer number $\ell$. To proceed further, we reformulate (35) as

$$\frac{M N_T}{N} |\varepsilon_{k_1} - \varepsilon_{k_2}| \neq |M\ell + k_2 - k_1| \tag{36}$$

and observe that the right-hand-side cannot be smaller than unity when $k_1 \neq k_2$. Hence, a *necessary* condition for code identifiability is that $M N_T |\varepsilon_{k_1} - \varepsilon_{k_2}| / N < 1$, which can be ensured by setting

$$|\varepsilon_k| < \frac{N}{2M N_T} \tag{37}$$

for $k = 1, 2, \dots, K$.

TABLE I

COMPUTATIONAL LOAD FOR EACH RANGING SUBCHANNEL

| Algorithm | Required Flops |
|-----------|----------------|
| MFE & MCD | $4M^2(QV + 18M) + 2MN_\epsilon(4M + 1)(M - K)$ |
| AHTE & AHPE | $4K^2M + 2KQV(4M + 6K + K^2 + 6)$ |
| AHCD | $4MQV(2K + 1)$ |

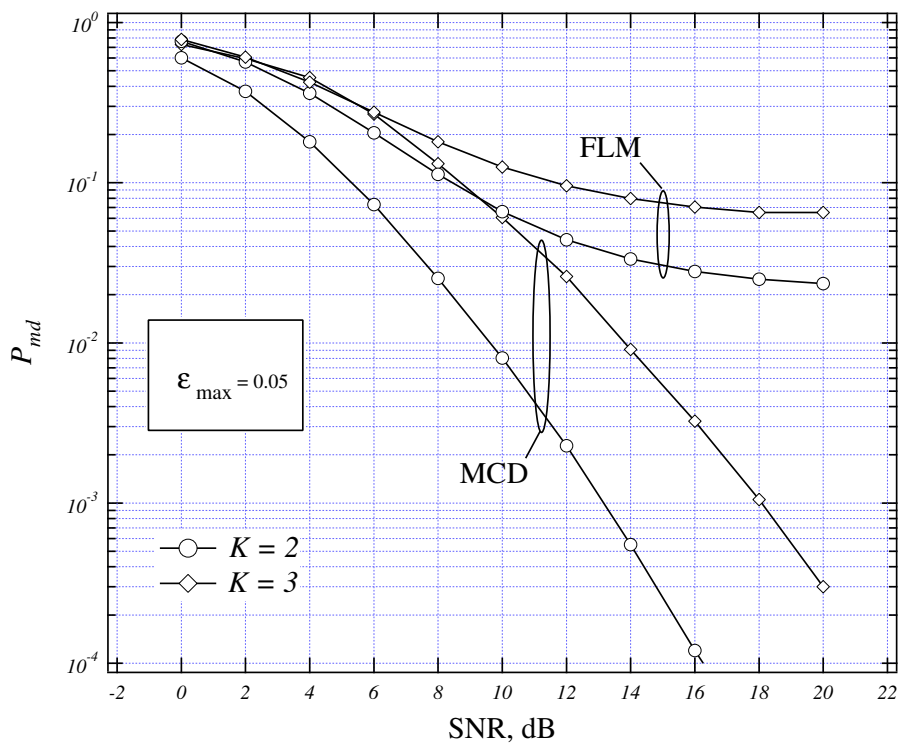

Fig. 1. Mis-detection probability vs. SNR with $K = 2$ or $3$ and $\varepsilon_{\max} = 0.05$.

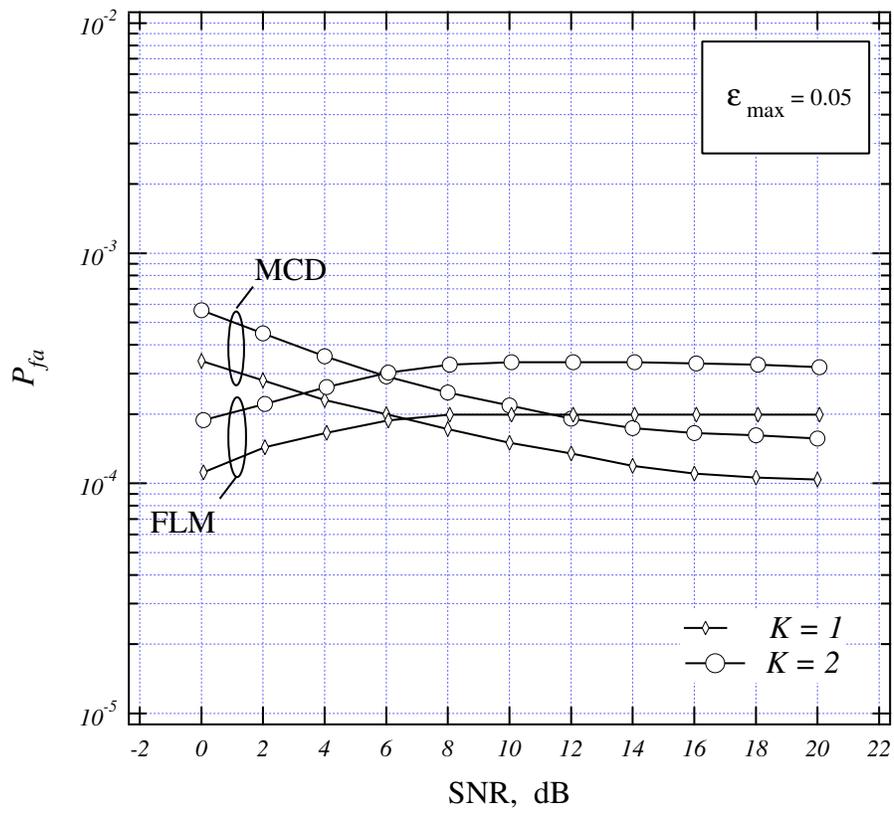

Fig. 2.   False alarm probability vs. SNR with $K = 1$ or $2$ and $\varepsilon_{\max} = 0.05$.

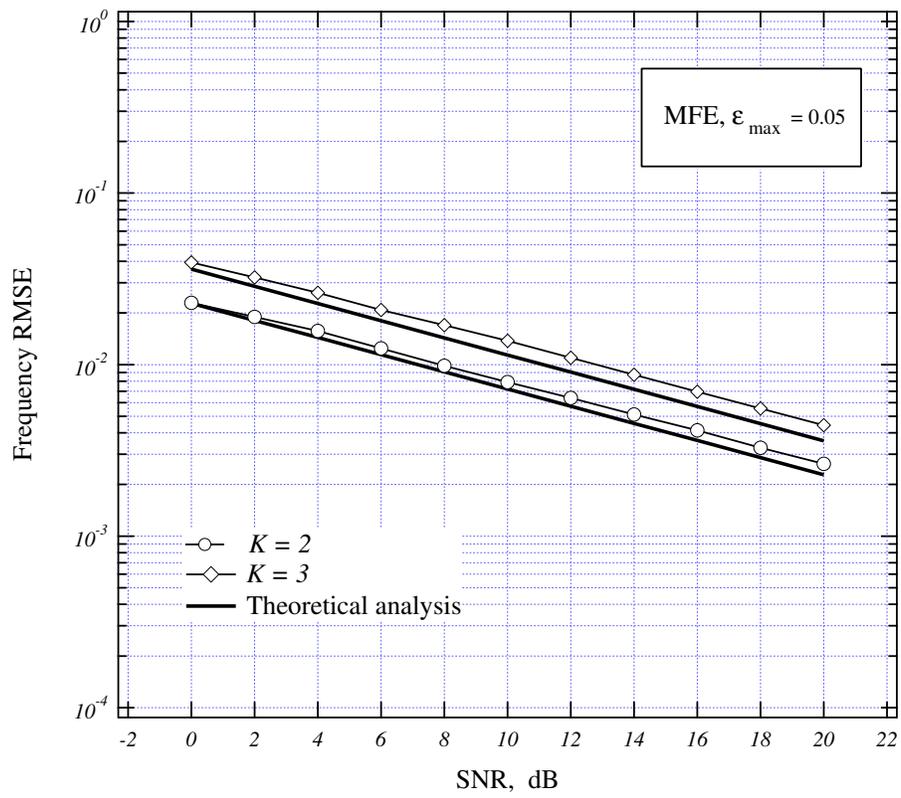

Fig. 3.   Frequency RMSE vs. SNR with $K = 2$ or $3$ and $\varepsilon_{\max} = 0.05$.

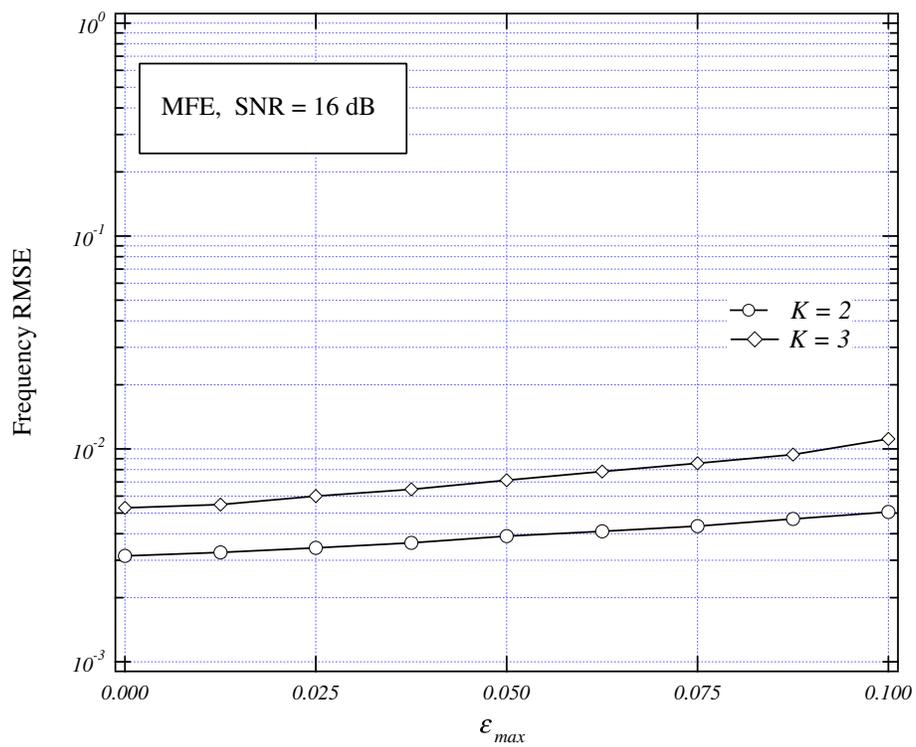

Fig. 4.  Frequency RMSE vs. $\varepsilon_{\max}$ with $K = 2$ or 3 and SNR= 16 dB.

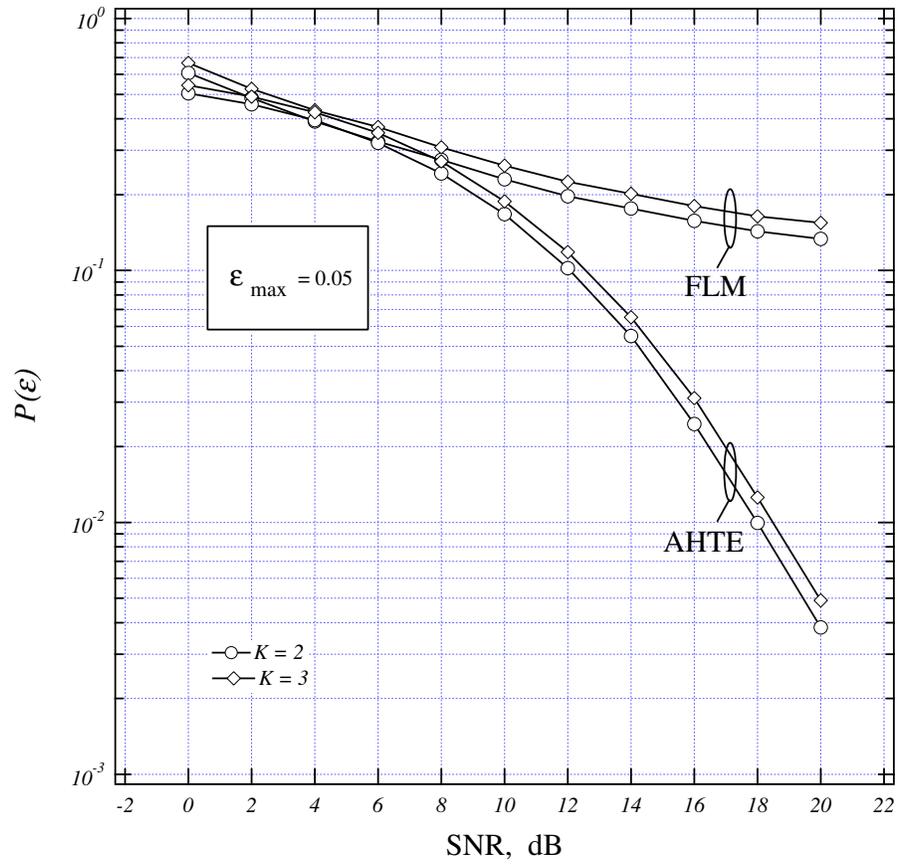

Fig. 5. $P(\epsilon)$ vs. SNR with $K = 2$ or 3 and $\varepsilon_{\max} = 0.05$.

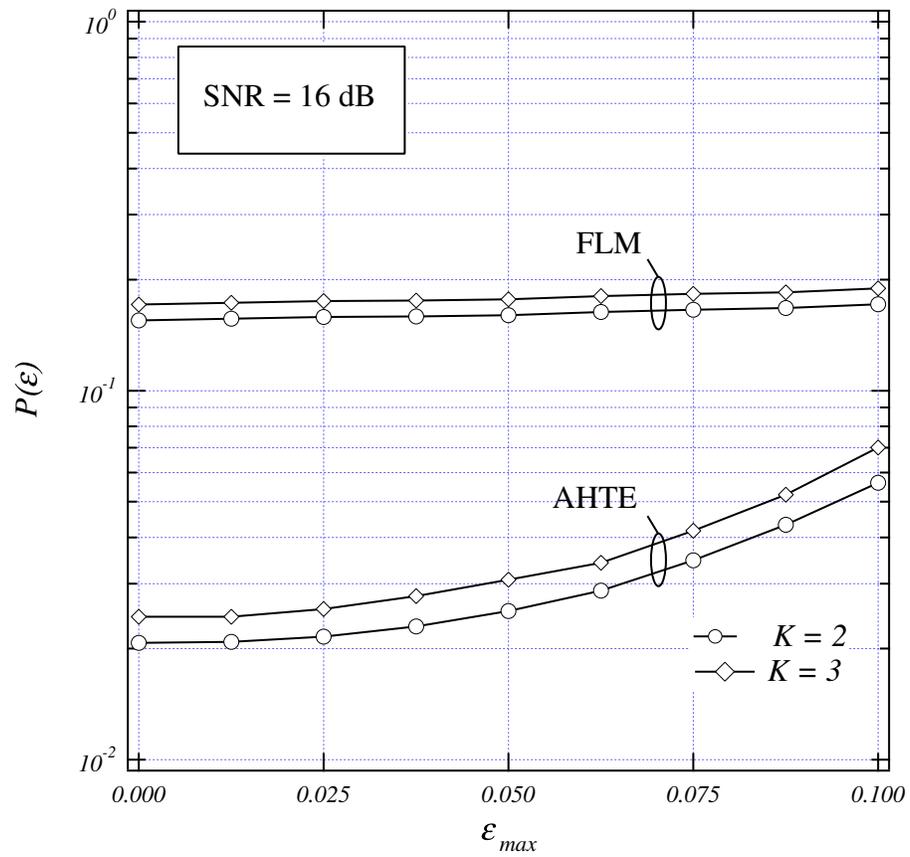

Fig. 6. $P(\epsilon)$ vs. $\varepsilon_{\max}$ with $K = 2$ or 3 and SNR= 16 dB.

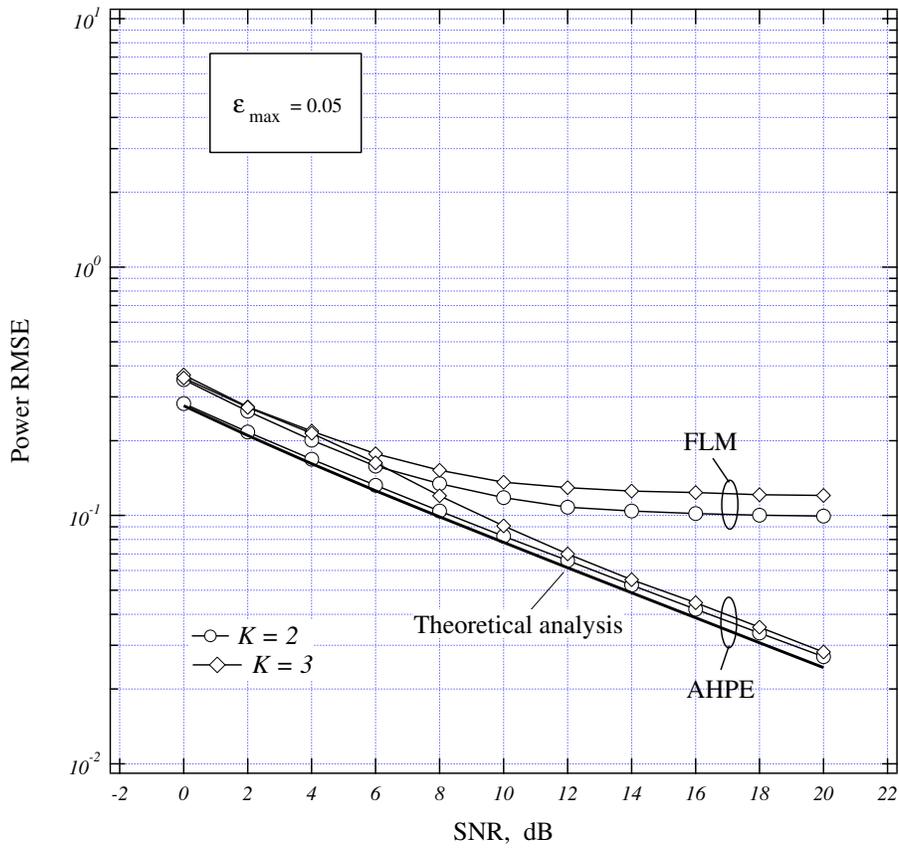

Fig. 7. RMSE of the power estimates vs. SNR with $K = 2$ or $3$ and $\varepsilon_{\max} = 0.05$.

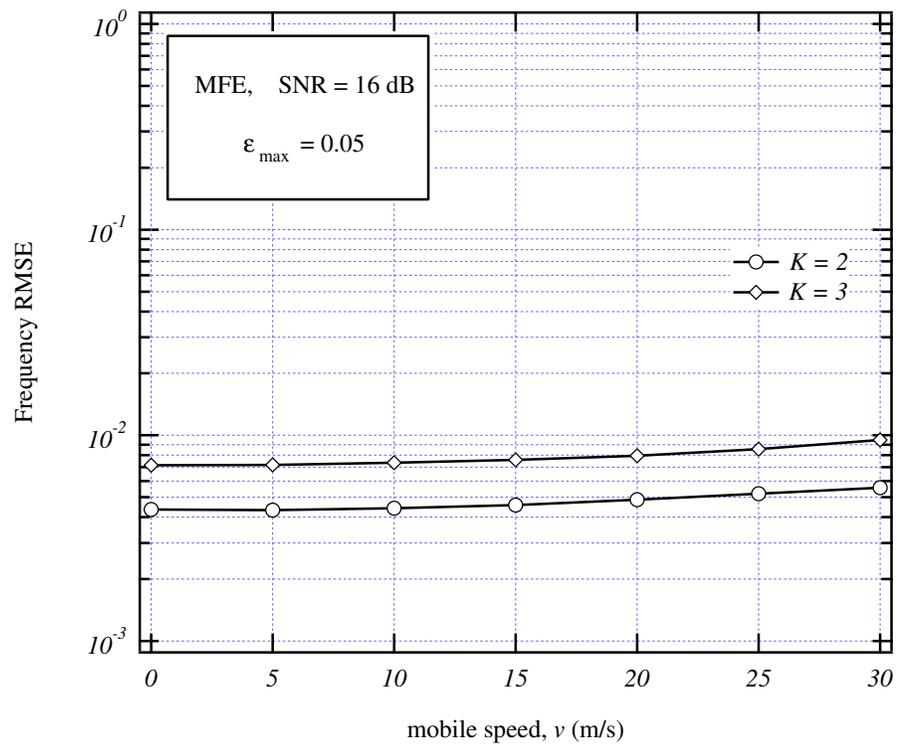

Fig. 8.   Frequency RMSE vs. *v* with *K* = 2 or 3 and SNR = 16 dB.

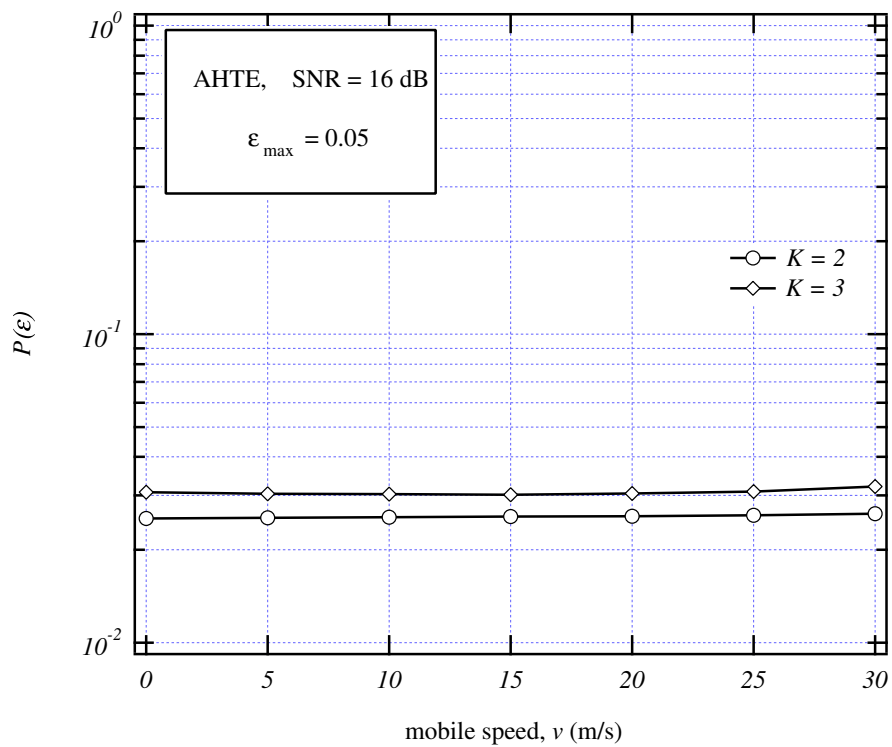

Fig. 9.  $P(\epsilon)$ vs. $v$ with $K = 2$ or 3 and SNR = 16 dB.

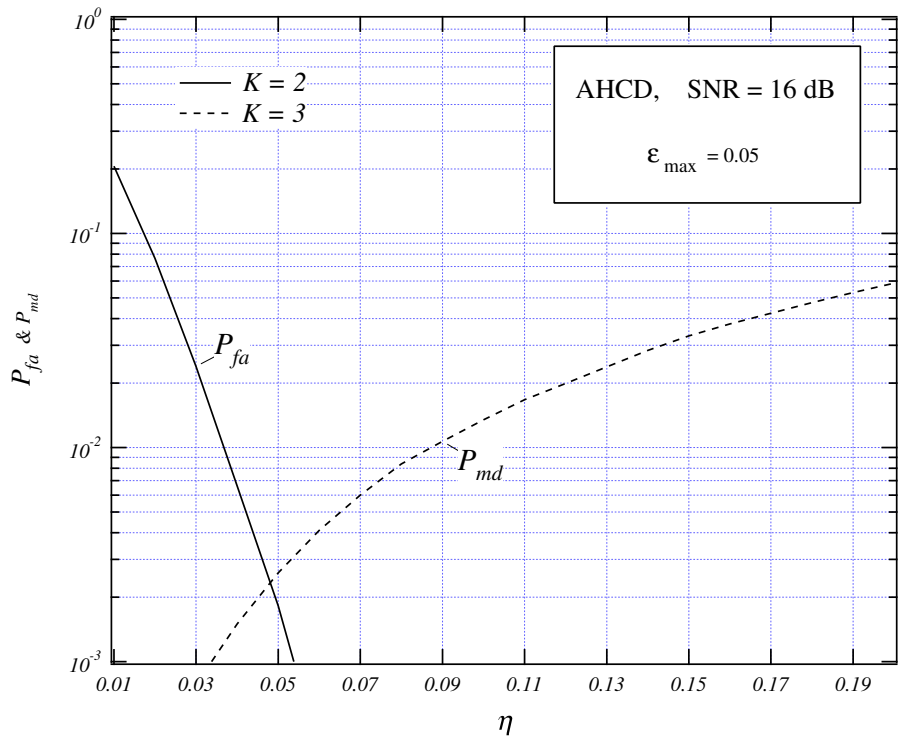

Fig. 10.   Performance of AHCD with SNR = 16 dB.